# Data-Driven Assisted Chance-Constrained Energy and Reserve Scheduling with Wind Curtailment

Xingyu Lei, *Student Member*, *IEEE*, Zhifang Yang, *Member*, *IEEE*, Junbo Zhao, *Senior Member*, *IEEE*, Juan Yu, *Senior Member*, *IEEE*

*Abstract*—Chance-constrained optimization (CCO) has been widely used for uncertainty management in power system operation. With the prevalence of wind energy, it becomes possible to consider the wind curtailment as a dispatch variable in CCO. However, the wind curtailment will cause impulse for the uncertainty distribution, yielding challenges for the chance constraints modeling. To deal with that, a data-driven framework is developed. By modeling the wind curtailment as a cap enforced on the wind power output, the proposed framework constructs a Gaussian process (GP) surrogate to describe the relationship between wind curtailment and the chance constraints. This allows us to reformulate the CCO with wind curtailment as a mixed-integer second-order cone programming (MI-SOCP) problem. An error correction strategy is developed by solving a convex linear programming (LP) to improve the modeling accuracy. Case studies performed on the PJM 5-bus and IEEE 118-bus systems demonstrate that the proposed method is capable of accurately accounting the influence of wind curtailment dispatch in CCO.

## I. INTRODUCTION

### A. Motivations

Uncertainties from high penetration of wind power can substantially affect the operation and reserve planning, power delivery of the transmission grid, and cost of total generation, etc. Recently, a wide variety of stochastic optimization methods is introduced to cope with wind uncertainty, including scenario-based approaches, robust optimization, and *chance-constrained optimization* (CCO) [1]. Among them, CCO ensures that the violation of uncertain constraints is restricted within a preset risk level. Also, CCO can be reformulated as a tractable convex program, which shows unique advantages in market scheduling and pricing.

Despite those advantages, however, it is challenging for existing CCO methods to consider wind curtailment. As the share of wind power rapidly increases, accommodating all wind power is a challenging task and considering wind curtailment in the dispatch model is needed. In fact, recent results shown in [2]-[4] suggest economically scheduling an acceptable level of wind curtailment rather than incorporating all wind power regardless of the operating expense. Hence, it is crucial to consider the wind curtailment as a dispatch variable in CCO, while maintaining the tractable modeling properties desired for market scheduling and pricing.

Existing CCO formulations typically assume that the uncertainty distributions have some desired properties, such as symmetry and unimodality. Based on that, the chance constraints can be reformulated as an analytical form. The key difficulty for the CCO with wind curtailment is that those properties of uncertainty distributions are hard to hold for true.

Specifically, the wind uncertainty distribution will become a truncated distribution with impulse, making existing formulations intractable. This paper proposes a data-driven framework to address this challenge. Procedures for ensuring the modeling accuracy of the chance constraints are also presented.

### B. Literature Review

CCO is an important tool for decision making in uncertain environments. It searches for the decision that satisfies all model constraints with a given probability while minimizing the objective function. There are many studies on solving *optimal power flow* (OPF) [5], [6], *unit commitment* (UC) problem [7], [8], and reserve scheduling [9] via CCO. The computationally tractable uncertainty representation aligns well with current industry practice [10], [11]. Besides, CCO internalizes the uncertainty of renewable resources in the market-clearing process using statistical moments of the underlying distribution, e.g, mean and standard deviation [12]-[14]. This unique characteristic of CCO provides a promising way to price the uncertainty of renewable resources in a chance-constrained stochastic electricity market. With some specific assumptions, CCO can express the influence of uncertain power injections on the system operational constraints, such as transmission line, and reserve requirement constants as a linear combination of statistical moments of uncertain variable distributions. By incorporating the stochastic optimization problem into a CCO formulation, the modeling convexity is maintained, which guarantees the convergence of scheduling while providing a unique dual variable for pricing.

To account for uncertainty in CCO and reformulate the chance constraint with analytic forms, two main reformulation approaches have been developed, namely the distribution-based and the scenario-based ones. The former one assumes a particular distribution of renewable output. For instance, [7], [15] assume that the uncertainty distributions of renewables are Gaussian while a Weibull distribution is used in [16]. However, the probability distribution of uncertain variables is typically challenging to model or even not known in practice due to nonstationarity [17]. The inclusion of wind curtailment further increases the complexity of uncertainty distribution. The scenario-based reformulation does not need the knowledge of the underlying distribution, and instead, tends to represent chance constraints by a given number of scenarios [18], [19]. However, it might lead to a high computational burden and lose the unique advantages in market scheduling and pricing. In this paper, the distribution-based reformulation is used and a data-driven framework is proposed to model the chance constraints analytically considering the wind curtailment. The convexity of the scheduling model is also preserved.

Proper wind curtailment allows a higher degree of system flexibility. To achieve that, a flexible dispatch margin of wind power is proposed in [20]. Wind generators are scheduled in the hour-ahead energy market while holding certain expected output as reserves. In [21], by utilizing the scenario-based stochastic methodology, an energy and reserve scheduling method is proposed considering wind power dispatch, and the flexibility of wind power is discussed. However, the modeling of chance constraints with wind curtailment is not studied to the best of our knowledge.

*C. Contribution*

In this paper, we propose a chance-constrained energy and reserve scheduling method considering wind curtailment. The main contributions are summarized as follows:
1. A GP-based data-driven reformulation of CCO is proposed for the chance constraint modeling considering wind curtailment. The GP surrogate is decomposed into two stages: the statistical-moment-based (i.e., the mean and standard deviation) rough approximation and the error compensation to reduce the learning complexity. Thus, the proposed framework can achieve an accurate reformulation while retaining useful statistical moments information. Note that no prior distributions of uncertain variables are needed in the formulation, which distinguishes with existing distribution-based methods.
2. A *chance-constrained energy and reserve co-optimization* (CC-ERCO) model is developed with the wind curtailment cap as a dispatch variable. The CC-ERCO model is also reformulated as a *mixed-integer second-order cone programming* (MI-SOCP) problem. To further enhance the modeling accuracy, a correction strategy is presented by solving a convex *linear programming* (LP) problem.

The rest of the paper is organized as follows. Section II shows the problem formulation and the data-driven chance constraints modeling is developed in Section III. Section IV shows the CCO reformulation and error correction strategy. Numerical results are analyzed in Section V and finally Section VI concludes the paper.

## II. PROBLEM FORMULATION

*A. CCO Considering Wind Curtailment*

In this subsection, the chance-constrained energy and reserve co-optimization model is developed, where the wind curtailment cap is taken as a dispatch variable to describe the correlation between wind uncertainty and wind curtailment. The wind power is modeled as the sum of forecasted output and the fluctuation caused by wind uncertainty:

$$W = W_{fc} + \Delta W(wc), \quad (1)$$

where $W$ is the actual wind power output; $W_{fc}$ and $\Delta W(wc)$ are the forecasted value and its uncertainty. The latter is described as a function of wind curtailment, as shown below:

$$\Delta W(wc) = \min\{\Delta W, wc - W_{fc}\}, \quad (2)$$

where $\Delta W$ is the uncertain part of wind power output that is assumed to be spatially uncorrelated; $wc$ represents the wind curtailment cap, which makes the uncertainty distribution of wind output a truncated one with impulse.

The dispatch model is shown as follows, where each thermal unit provides three services: the energy service, the up spinning reserve service, and the down spinning reserve service. Formally, we have

$$\min \sum_{i \in I_G} \left[ E(c_{e,i} P_i) + c_{r,i}(R_{up,i} + R_{dn,i}) \right] \quad (3a)$$

$$s.t. \quad \sum_{i \in I} (P_{sc,i} + W_{fc,i} - D_i) = 0 \quad (3b)$$

$$\Pr(\sum_{i \in I} PTDF_{li} \times (P_i + W_i - D_i) \leq F_l) \geq 1 - \varepsilon \quad (3c)$$

$$\Pr(-\sum_{i \in I} PTDF_{li} \times (P_i + W_i - D_i) \leq F_l) \geq 1 - \varepsilon \quad (3d)$$

$$\Pr(\Delta P_i(\boldsymbol{wc}) \leq R_{up,i}) \geq 1 - \varepsilon \quad (3e)$$

$$\Pr(\Delta P_i(\boldsymbol{wc}) \geq -R_{dn,i}) \geq 1 - \varepsilon \quad (3f)$$

$$P_{sc,i} + R_{up,i} \leq P_{\max,i} \quad (3g)$$

$$P_{sc,i} - R_{dn,i} \geq P_{\min,i} \quad (3h)$$

$$W_{fc,i} \leq wc_i \leq W_{\max,i} \quad (3i)$$

$$P_i = P_{sc,i} + \Delta P_i(\boldsymbol{wc}) \quad (3j)$$

$$W_i = W_{fc,i} + \Delta W_i(wc_i) \quad (3k)$$

$$\Delta P_i(\boldsymbol{wc}) = -\beta_i \sum_{i \in I} \Delta W_i(wc_i). \quad (3l)$$

In this model, $c_{e,i}$ and $c_{r,i}$ are generator costs for production and reserve capacity, respectively; $P_i$, $P_{sc,i}$, $\Delta P_i(\boldsymbol{wc})$ are actual generation output, scheduled generation output, and generation response to balance the system uncertainty, respectively; $\boldsymbol{wc}$ is the vector of wind curtailment caps; $D_i$ represents demand quantity; $R_{up,i}$ and $R_{dn,i}$ are up and down reserves of generators; $i$ and $l$ are the indices of bus and transmission lines; $PTDF_{li}$ represent the power transfer distribution factor of bus $i$ to line $l$; $\varepsilon$ is the confidence level in the chance constraints; $I$, $I_G$, and $I_W$ are the sets of buses, thermal generators, and wind powers, respectively.

The total offer cost in (3a) includes the expected energy cost and the offer costs for up/down-spinning reserves. The system-wide power balance is enforced in (3b), which balances the total output of conventional and wind power generations and demand on the forecasted operating point. This constraint ensures a feasible solution for the forecasted system state [22]. Constraints (3c)-(3f) are the chance constraints considering the impact of the wind uncertainty on the transmission overloading and reserve requirement. Constraint (3l) enforces that the total generation response equals the wind uncertainty. Note that the participation factors $\beta_i$ can be chosen in different ways. In this paper, we assume that each generator contributes according to its maximum output [23], [24]. The expression of $\beta_i$ is given as follows:

$$\beta_i = \frac{P_{\max,i}}{\sum_{j \in I_G} P_{\max,j}} \quad \forall i \in G. \quad (4)$$

*B. Problem Statement*

The chance constraints (3c)-(3f) restrict the probability of constraint violation under wind power uncertainty. In traditional methods, the mean $\mu_i$ and the standard deviation $\sigma_i$ of the wind uncertainty are assumed to be known constants [8], [14], [22]. Hence, taking the transmission constraint (5c) as an example, the chance constraint can be formulated as follows:

$$\sum_{i \in I} PTDF_{li}(P_{sc,i} + W_{fc,i} - D_i) + \sum_{i \in I_W} K_{li}\mu_i + \Gamma_\varepsilon \sqrt{\sum_{i \in I_W} K_{li}^2 \sigma_i^2} \leq F_l, \quad (5)$$

where $\Gamma_\varepsilon$ is a constant that represents coefficient on the standard deviation component; $K_{li}$ is the sensitivity factor for the power flow of transmission line $l$ with respect to the fluctuation of wind power plant $i$, which is given by

$$K_{li} = -\sum_{k \in I} PTDF_{lk}\beta_k + PTDF_{li}. \quad (6)$$

Defining $PF_{sc,l} = \sum_{i \in I} PTDF_{li}(P_{sc,i} + W_{fc,i} - D_i)$, (5) can be expressed as:

$$PF_{sc,l} + \sum_{i \in I_W} K_{li}\mu_i + \Gamma_\varepsilon \sqrt{\sum_{i \in I_W} K_{li}^2 \sigma_i^2} \leq F_l. \quad (7)$$

Considering wind curtailment dispatch, the reformulation expressed in (7) becomes (8) because the mean and deviation of wind is a function of curtailment cap:

$$PF_{sc,l} + \sum_{i \in I_W} K_{li}\mu_i(wc_i) + \Gamma_\varepsilon(\boldsymbol{wc}) \sqrt{\sum_{i \in I_W} K_{li}^2 \sigma_i(wc_i)^2} \leq F_l. \quad (8)$$

The chance constraint (8) may be intractable for three reasons. First, probability distributions are not known in practice or contain large errors. Second, because the wind curtailment cap is a dispatch variable in our proposed model, the wind uncertainty distribution will change along with the optimization procedure. It makes the statistical moments of wind uncertainty ($\mu_i$, $\sigma_i$) and $\Gamma_\varepsilon$ a function of $wc_i$, and the analytic expression is difficult to obtain. Third, the tight coupling between $\Gamma_\varepsilon(\boldsymbol{wc})$ and $\sqrt{\sum_{i \in I_W} K_{li}^2 \sigma_i(wc_i)^2}$ makes constraint (8) complicated.

### III. Data-Driven Reformulation of Chance Constraints

To reformulate chance constraints into a tractable convex form, a data-driven framework is proposed. It has two main blocks: 1) a rough approximation and 2) the error compensation.

#### A. Basic Idea

To reduce the modeling complexity, we decouple $\Gamma_\varepsilon(\boldsymbol{wc})$ and $\sqrt{\sum_{i \in I_W} K_{li}^2 \sigma_i(wc_i)^2}$ by the Taylor series expansion of $\Gamma_\varepsilon(\boldsymbol{wc})$, yielding

$$PF_{sc,l} + \sum_{i \in I_W} K_{li}\mu_i(wc_i) + (\Gamma_{\varepsilon 0} + \sum_{i=1}^{\infty} C_i(\boldsymbol{wc})) \sqrt{\sum_{i \in I_W} K_{li}^2 \sigma_i(wc_i)^2} \leq F_l, \quad (9)$$

where $C_i(\boldsymbol{wc})$ is the $i$th-order of the Taylor series. Defining

$$\kappa_{\max,l}(\boldsymbol{wc}) = \sum_i C_i(\boldsymbol{wc}) \sqrt{\sum_{i \in I_W} K_{li}^2 \sigma_i(wc_i)^2}, \quad (9) \text{ can be expressed as:}$$

$$PF_{sc,l} + \sum_{i \in I_W} K_{li}\mu_i(wc_i) + \Gamma_{\varepsilon 0} \sqrt{\sum_{i \in I_W} K_{li}^2 \sigma_i(wc_i)^2} + \kappa_{\max,l}(\boldsymbol{wc}) \leq F_l. \quad (10)$$

Then, the GP surrogates are constructed to obtain the analytic expression of $\mu_i(wc_i)$, $\sigma_i(wc_i)$, and $\kappa_{\max,l}(\boldsymbol{wc})$, which will be shown later. To elaborate on our idea, we define the uncertainty margin of a chance constraint as a tightening boundary of the chance constraint, which is necessary to secure the system against uncertainty [22]. The uncertainty margin of constraint (10) can be expressed as follows:

$$PF_{sc,l} + UM_{\max,l}(\boldsymbol{wc}) \leq F_l,$$

$$UM_{\max,l}(\boldsymbol{wc}) = \underbrace{\sum_{i \in I_W} K_{li}\mu_i(wc_i) + \Gamma_{\varepsilon 0} \sqrt{\sum_{i \in I_W} K_{li}^2 \sigma_i(wc_i)^2}}_{\text{Rough approximation}} + \underbrace{\kappa_{\max,l}(\boldsymbol{wc})}_{\text{Error compensation}}, \quad (11)$$

where the term $\left(\sum_{i \in I_W} K_{li}\mu_i(wc_i) + \Gamma_{\varepsilon 0} \sqrt{\sum_{i \in I_W} K_{li}^2 \sigma_i(wc_i)^2}\right)$ is similar to the expression of traditional chance constraint reformulation methods. They can be regarded as a rough approximation of the uncertainty margin. The auxiliary variable $\kappa_{\max,l}(\boldsymbol{wc})$ can be seen as the error compensation. The basic idea of the proposed data-driven reformulation framework is to construct GP surrogates for $\boldsymbol{wc}$ and $UM_{\max,l}(\boldsymbol{wc})$. To this end, we provide a rough approximation of uncertainty margin based on the statistical moments of wind uncertainty. Then, the error compensation is obtained by simulating the empirical uncertainty margin from historical data. The proposed framework is shown in Fig. 1. It should be noted that the uncertainty margin is only affected by the wind curtailment cap and the wind power uncertainty itself. This property allows us to develop a data-driven reformulation of chance constraints that can be done offline.

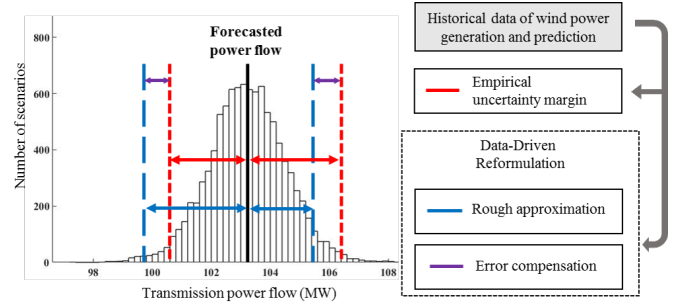

Fig. 1. Data-driven reformulation of chance constraints, where the transmission constraints are taken as an example.

#### B. Rough Approximation of Uncertainty Margin

In the rough approximation presented in (11), there exist several technical challenges. First, the standard deviation $\Gamma_{\varepsilon 0}$ is usually determined based on the information of specific fixed probability distribution. But in our case, the standard deviation $\Gamma_{\varepsilon 0}$ is hard to confirm, because the probability distribution of wind uncertainty is unknown and changes along with the optimization procedure. Second, the relationship between wind curtailment caps $wc_i$ and the statistical moments of the wind uncertainty (e.g., $\mu_i(wc_i)$, $\sigma_i(wc_i)$) need to be analytically expressed. To overcome them, the standard deviation $\Gamma_{\varepsilon 0}$ is determined based on the central limit theorem [26], [27]. Then, the analytical relationship between $wc_i$ and $\mu_i(wc_i)$, $\sigma_i(wc_i)$ is obtained by GP regression through historical data. The detailed procedures are shown below.

1) *Determination of Standard Deviation* $\Gamma_{\varepsilon 0}$

According to [14], [27], the standard deviation $\Gamma_{\varepsilon 0}$ is determined based on the knowledge of the random variable $\Delta W_i(wc_i)$. For example, if $\Delta W_i(wc_i)$ is Gaussian, $\Gamma_{\varepsilon 0}$ can be denoted as $\Psi^{-1}(1-\varepsilon)$, which is the inverse cumulative distribution function of a standard Gaussian distribution. However, in our case, the distribution function is unknown with the wind curtailment. Based on the central limit theorem, no matter how the wind curtailment cap $wc_i$ change, the distribution of $\Delta W_i(wc_i)$ is often close to a Gaussian in practice, making $\Psi^{-1}(1-\varepsilon)$ a good rough approximation to the real uncertainty margin. Hence, the standard deviation component $\Gamma_{\varepsilon 0}$ can be denoted as follows:

$$\Gamma_{\varepsilon 0} = \Psi^{-1}(1-\varepsilon) = \sqrt{(1-\varepsilon)/\varepsilon}. \quad (12)$$

2) *Regression for* $\mu_i(wc_i)$ *and* $\sigma_i(wc_i)$

To obtain the analytical relationship between $wc_i$ and $\mu_i(wc_i)$, $\sigma_i(wc_i)$, GP is introduced [28], [29]. Unlike other machine learning approaches base on deep neural networks [30], [31], GP can provide an analytical regression function, which is well suited to our problem. The regression function of GP can be seen as a linear combination of $K$ kernel functions, and each one is centered on a training point, which is shown as follows:

$$f(\boldsymbol{x}_*) = \sum_{k=1}^{K} \alpha_k C(\boldsymbol{x}_i, \boldsymbol{x}_*), \quad (13)$$

where $f(\boldsymbol{x}_*)$ is the output of GP ($\mu_i(wc_i)$ and $\sigma_i(wc_i)$ in our case); the input vector $\boldsymbol{x}_*$ is the wind curtailment cap $wc_i$; $\boldsymbol{x}_k$ is the input vector of training samples; $\alpha_k$ represents the weight parameter, which is obtained from training samples; $C(\cdot, \cdot)$ is the kernel function that represents the covariance function. Here, we choose *squared exponential* (SE) covariance kernel function for our regression problem, i.e.,

$$C_{SE}(\boldsymbol{x}_k, \boldsymbol{x}_*) = \tau^2 \exp(-\frac{(\boldsymbol{x}_k - \boldsymbol{x}_*)^T (\boldsymbol{x}_k - \boldsymbol{x}_*)}{2l^2}), \quad (14)$$

where $\tau$ and $l$ are hyper-parameters that can be obtained through the GP training process.

**Algorithm 1** Training samples generation for $\mu_i(wc_i), \sigma_i(wc_i)$

**Inputs:** $k=0$, $t_{wc}$ (step-length of enumeration method), $\{\xi^n\}_{n=1}^{N}$ (historical dataset), $wc_{i,0}$ (initial point of $wc_i$)

**repeat**

For each scenario $\xi^n$, obtain $\xi^n_{wc_{i,k}}$ by considering $wc_{i,k} = wc_{i,0} + kt_{wc}; \forall i \in I_W$

Obtain $\mu_i(wc_{i,k})$, $\sigma_i(wc_{i,k})$ by performing statistical analysis on dataset $\{\xi^n_{wc_{i,k}}\}_{n=1}^{N}$.

$k = k+1$.

**until** $wc_{i,k} > W_{\max,i}$

**Return:** Training sample set including input set $\{wc_{i,k}\}$ and output set $\{\mu_i(wc_{i,k})\}, \{\sigma_i(wc_{i,k})\}$.

To develop a trained GP, we obtain the training samples using the enumeration method described in Algorithm 1 for each wind farm from the historical dataset $\{\xi^n\}_{n=1}^{N}$, where $N$ is the total number of scenarios.

After Algorithm 1, the training samples of GP are obtained. For each wind farm, a well-trained GP is developed as a surrogate model for $wc_i$ and $\mu_i(wc_i)$, $\sigma_i(wc_i)$. Then, $\mu_i(wc_i)$, $\sigma_i(wc_i)$ can be analytically expressed as follows:

$$\mu_i(wc_i) = \sum_{k=1}^{K} \alpha_{\mu_i,k} C_{SE}(wc_k, wc_i)$$

$$= \sum_{k=1}^{K} \alpha_{\mu_i,k} \tau_{\mu_i}^2 \exp(-\frac{(wc_k - wc_i)^T (wc_k - wc_i)}{2l_{\mu_i}^2}),$$

$$\sigma_i(wc_i) = \sum_{k=1}^{K} \alpha_{\sigma_i,k} C_{SE}(wc_k, wc_i)$$

$$= \sum_{k=1}^{K} \alpha_{\sigma_i,k} \tau_{\sigma_i}^2 \exp(-\frac{(wc_k - wc_i)^T (wc_k - wc_i)}{2l_{\sigma_i}^2}), \quad (15)$$

where $\mu_i(wc_i)$ and $\sigma_i(wc_i)$ are non-linear functions because of the kernel function (14). Similar to [32], the piecewise linearization technique is used. Thus, based on the central limit theorem and GP, a rough approximation of the uncertainty margin is achieved.

*C. Error Compensation*

In this subsection, the error compensation is obtained by simulating the empirical uncertainty margin. Let us review the uncertainty margin defined in (11). $UM_{\max,l}$ can be seen as an approximation of the quantile of the distribution of transmission power flow caused by wind uncertainty. By sampling the wind uncertainty vector $\Delta \boldsymbol{W}(\boldsymbol{wc})$ from historical data and calculating the uncertainty part of power flows $\Delta PF_l$ caused by wind uncertainty, we can compute the empirical uncertainty margin $UM_{max,l}^{em}$[1]. For example, assuming that we want to ensure the violation of (3c) is restricted within a preset probability $\varepsilon$. $N$ scenarios are used to estimate $UM_{max,l}^{em}$. To compute the upper quantiles, we calculate $\Delta PF_l$ for each scenario $\xi^n$. The empirical uncertainty margin of (1d) $UM_{max,l}^{em}$ is defined as the upper quantile corresponding to the $(\varepsilon N)^{th}$ largest magnitude $\Delta PF_l^{\varepsilon N}$ [2]. Then, the correction variable $\kappa_{\max,l}(\boldsymbol{wc})$ can be calculated by

$$\kappa_{\max,l}(\boldsymbol{wc}) = UM_{\max,l}^{em}(\boldsymbol{wc}) - \sum_{i \in I_*} K_{il} \mu_i(wc_i) - \Gamma_{\varepsilon 0} \sqrt{\sum_{i \in I_*} K_{li}^2 \sigma_i(wc_i)^2}. \quad (16)$$

Note that the analytical formulation of the correction variable $\kappa_{\max,l}(\boldsymbol{wc})$ is also unknown. To obtain the analytical relationship between $\boldsymbol{wc}$ and $\kappa_{\max,l}(\boldsymbol{wc})$, we build a surrogate model for $\kappa_{\max,l}(\boldsymbol{wc})$ using GP. Different from Section III.B, the linear kernel function $C_{LIN}(\cdot, \cdot)$ is introduced to maintain the convexity of CCO. The challenge is to obtain GP training samples. To this end, Algorithm 2 is proposed. The goal of it is to obtain the error compensation. For each chance constraint, the error variable is different. This work can be done offline, which does not affect the efficiency of the CCO. The expression of $\kappa_{\max,l}(\boldsymbol{wc})$ is shown as follows:

---

[1] The expressions of uncertainty part of power flows and reserve requirements in (3c)-(3f) are provided in Appendix A.

[2] For chance constraint (3b), the empirical uncertainty margin is defined as the lower quantile corresponding to the $(\varepsilon N)^{th}$ lowest magnitude $\Delta PF_l^{(1-\varepsilon)N}$.

$$\kappa_{\max,l}(\boldsymbol{wc}) = \sum_{k=1}^{K} \alpha_{\max,l,k} C_{LIN}(\boldsymbol{wc}_k, \boldsymbol{wc})$$
$$= \sum_{k=1}^{K} \alpha_{\max,l,k} \frac{\boldsymbol{wc}_k^T \boldsymbol{wc}}{l_{LIN}^2}, \quad (17)$$

where $l_{LIN}$ is the hyper-parameter of the linear kernel function $C_{LIN}(\cdot,\cdot)$. Due to the limited learning ability of the linear kernel, the GP surrogate of the error variable presented in (17) might be inaccurate. To address that, we further develop a correction strategy in the next section to guarantee accuracy.

**Algorithm 2** Training sample generation for $\kappa_{\max,l}(\boldsymbol{wc})$

**Inputs:** $k=0$, $t_{wc,i}$; $\forall i \in I_W$, $\{\xi^n\}_{n=1}^N$, $wc_{i,0}$

  **repeat**

  For each scenario $\xi^n$, obtain $\xi^n_{wc_{i,k}}$ by considering $wc_{i,k} = wc_{i,0} + kt_{wc,i}$; $\forall i \in I_W$

  Obtain $UM^{em}_{max,l}(\boldsymbol{wc}_k)$ by calculating the uncertainty part of power flows for each $\xi^n_{wc_{i,k}}$

  Obtain $\kappa_{\max,l}(\boldsymbol{wc}_k)$ base on Eq.(16).

  $k = k+1$;

  **until** $wc_{i,k} > W_{\max,i}$;

**Return:** Training sample set including input set $\{wc_{i,k}\}$ and output set $\{\kappa_{\max,l}(\boldsymbol{wc}_k)\}$.

To summarize, a data-driven reformulation framework for chance constraints is proposed without the knowledge of uncertainty distribution of wind output. Two stages are developed to make an accurate approximation of uncertainty margin, i.e., rough approximation and error compensation. In fact, considering (16) and the discussion presented above, the empirical uncertain margin $UM^{em}_{max,l}(\boldsymbol{wc})$ in (16) can be seen as a function of $\boldsymbol{wc}$. Naturally, there might exist another way to solve the reformulation problem by direct learning its relationship. The reason for developing the two-stage framework is two-fold. First, the relationship between $\boldsymbol{wc}$ and $UM^{em}_{max,l}(\boldsymbol{wc})$ is rather complex. Direct learning occasionally leads to large errors. By contrast, the proposed framework reduces the learning complexity of GP by including physical characteristics of the power grid (e.g., $K_{li}$) while enhancing the interpretability. Second, the statistical information of wind uncertainty is retained, which provides intuitive information for system operators and power consumers to manage wind uncertainty. The overall flowchart is shown in Fig. 2.

IV. CC-ERCO MODEL REFORMATION AND CORRECTION STRATEGY

With the data-driven reformulation in Section III, we now reformulate the CC-ERCO model (1) as an MI-SOCP problem. Then a correction strategy is presented to ensure the modeling accuracy. After implementing the proposed reformulation framework with piecewise linearization technique, the CC-ERCO model is formulated as:

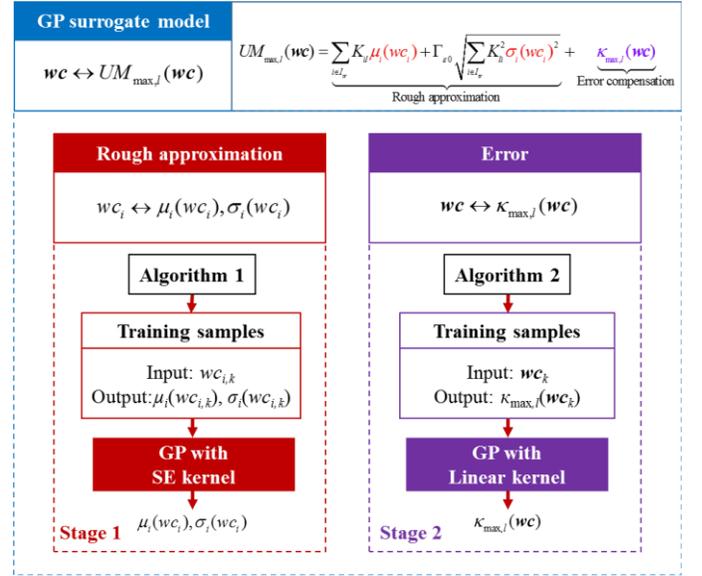

Fig. 2. Flowchart of the proposed two-stage data-driven framework.

$$\min \sum_{i \in I} \left( c_i^1 \cdot (P_{sc,i} - \beta_i \sum_{i \in I} \mu_i(wc_i)) + c_i^2 \cdot (R_{up,i} + R_{dn,i}) \right) \quad (18a)$$

s.t. Constraints (3b) and (3g)-(3i);

$$PF_{sc,l} + \sum_{i \in I_w} K_{li} \mu_i(wc_i) + \Psi^{-1}(1-\varepsilon)\Lambda_{PF,l} + \kappa_{\max,l}(\boldsymbol{wc}) \leq F_l \quad (18b)$$

$$-PF_{sc,l} - \sum_{i \in I_w} K_{li} \mu_i(wc_i) + \Psi^{-1}(1-\varepsilon)\Lambda_{PF,l} + \kappa_{\min,l}(\boldsymbol{wc}) \leq F_l \quad (18c)$$

$$-\beta_i \sum_{i \in I} \mu_i(wc_i) + \Psi^{-1}(1-\varepsilon)\Lambda_{g,i} + \kappa_{up,i}(\boldsymbol{wc}) \leq R_{up,i} \quad (18d)$$

$$\beta_i \sum_{i \in I} \mu_i(wc_i) + \Psi^{-1}(1-\varepsilon)\Lambda_{g,i} + \kappa_{dn,i}(\boldsymbol{wc}) \leq R_{dn,i} \quad (18e)$$

$$\sqrt{\sum_{i \in I_w} K_{li}^2 \sigma_i(wc_i)^2} \leq \Lambda_{PF,l} \quad (18f)$$

$$\sqrt{\beta_i^2 \sum_{i \in I_w} \sigma_i(wc_i)^2} \leq \Lambda_{g,i}, \quad (18g)$$

where $\kappa_{\min,i}(\boldsymbol{wc})$, $\kappa_{up,i}(\boldsymbol{wc})$ and $\kappa_{dn,i}(\boldsymbol{wc})$ are the ancillary variables of constraints (3d)-(3f), respectively; $\Lambda_{PF,l}$ and $\Lambda_{r,l}$ are the slack variables for the standard deviation of transmission power flow and reserve requirement, respectively.

The model formulated in (18) is a mixed-integer second-order cone programming (MI-SOCP) problem that can be solved using off-the-shelf mathematical programming solvers.

To correct the linearization errors caused by piecewise linearization and linear kernel function, a correction strategy is presented, which has four steps.

1) Step 1: After solving the MI-SOCP problem, we fix the optimal wind curtailment dispatch variables $\widehat{\boldsymbol{wc}}$.

2) Step 2: The error variable $\kappa^{em}(\boldsymbol{wc})$ is learned by the SE covariance function presented in (14).

3) Step 3: By replacing the wind curtailment dispatch correlated variables with constants $\mu_i(\widehat{wc}_i)$, $\sigma_i(\widehat{wc}_i)$ and $\kappa^{em}(\widehat{wc})$ the MI-SOCP problem is converted into a LP problem, as shown in Appendix B.

4) Step 4: By solving the LP problem, the modeling accuracy is ensured through the generator adjustment and the optimal solutions $P_{sc,i}^*$, $wc_i^*$, $R_{up,i}^*$, $R_{dn,i}^*$ are obtained.

## V. NUMERICAL RESULTS

In this section, the proposed method is tested on the PJM 5-bus system to illustrate its effectiveness while the IEEE 118-bus system is used to demonstrate its scalability to larger system. All simulations are performed on a PC equipped with Intel Core i5-4460 CPU @ 3.20GHz 16GB RAM. The algorithm is implemented in MATLAB, using Gurobi 9.0.2 to solve the proposed MIQCP and LP.

The following methods are compared:
**M0: Proposed method.**
M1: Traditional reformulation method [14], [25].

### A. An Illustrative Example on PJM 5-Bus System

The test system is modified from the original PJM 5-bus system by connecting a wind power plant at bus 2. The system is depicted in Fig. 3. The forecasted wind power output is 200 MW and its uncertainty is assumed to follow a Gaussian distribution, where the mean and standard deviation are 0 MW and 200 MW, respectively. The probability $\varepsilon$ in the chance constraints is 5%.

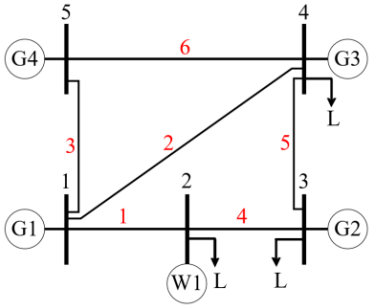

Fig. 3. PJM 5-bus system.

### B. Accuracy of Chance Constraint Reformulation

#### i) Performance without wind curtailment

We first show the proposed method achieves better performance than the traditional method even without wind curtailment. We set the wind curtailment dispatch variable to infinity in the proposed method and compare it with M1. The results are shown in Table I.

TABLE I
ACCURACY EVALUATION IN PJM 5-BUS SYSTEM ( $WC=+\infty$ )

| Methods | $M0(wc=+\infty)$ | M1 |
|---|---|---|
| Total operational cost ($) | **1.712×10⁴** | 1.714×10⁴ |
| Energy cost ($) | **1.385×10⁴** | 1.386×10⁴ |
| Reserve cost ($) | **3.274×10³** | 3.290×10³ |
| Total up reserve capacity (MW) | **326.86** | 329.00 |
| Total down reserve capacity (MW) | **328.03** | 329.00 |
| Max transmission violation | **4.93%** | 4.89% |
| Max generation violation | **5.00%** | 4.95% |

The results show that the proposed method provides a similar optimal solution as compared with the traditional method but maintains a more strict constraint violation probability. The proposed method can be used to model CCO without wind curtailment.

#### ii) Performance with wind curtailment

Considering wind curtailment, the traditional reformulation method might be inaccurate because the desired properties of wind uncertainty distributions (e.g., symmetry and unimodality) are disturbed by wind curtailment. By contrast, the proposed method can dispatch the wind curtailment while ensuring the modeling accuracy. To demonstrate that, for M1, we set the wind curtailment cap as $wc_i^*$ which is the optimal solution obtained by the proposed method. Then, the constants $\mu_i(wc_i)$ and $\sigma_i(wc_i)$ are calculated by performing statistical analysis. The max constraint violation probabilities of M0 and M1 in the PJM 5-bus system are obtained by running a Monte-Carlo simulation (10000 samples), as shown in Table II. For each sample, the actual realization of the transmission power flow and the generator power output considering the wind uncertainty are depicted in Fig. 4 and Fig. 5 for M0 and M1.

TABLE II
ACCURACY EVALUATION IN PJM 5-BUS SYSTEM

| Methods | **M0** | M1($wc=wc_i^*$) |
|---|---|---|
| Max transmission violation | **5.00%** | 6.88% |
| Max generation violation | **5.00%** | 6.88% |

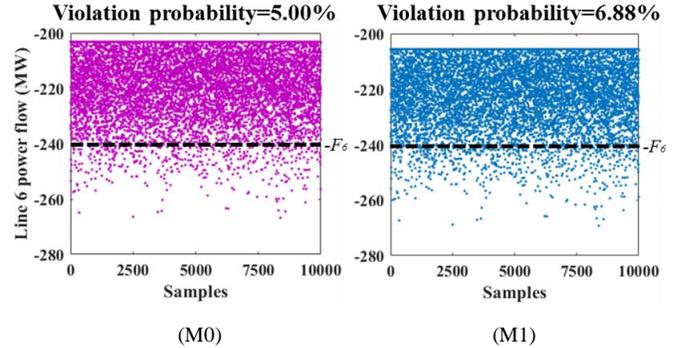

Fig. 4. Transmission line 6 power flow of 10000 samples.

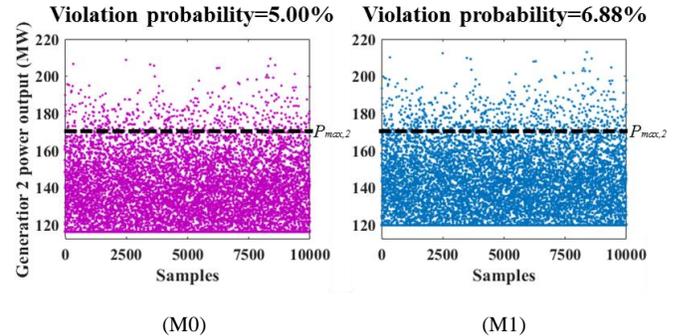

Fig. 5. Generator 2 power output of 10000 samples.

From Table I, it is clear that the traditional reformulation method is intractable considering wind curtailment because the wind uncertainty distribution is a truncated distribution with impulse. The proposed data-driven framework reformulates the chance constraints precisely. From Fig. 4 and Fig. 5, it is obvious that both the transmission power flow and generation output change significantly under the uncertainty environment. Note that the limit of Line 6 is 240 MW, and the Generator 2 output limit is 170MW. Compared with the traditional method, the constraint violation probabilities are significantly improved.

### C. Economic Performance of the Proposed Method

To further demonstrate the economic efficiency of the proposed method, the operational cost and total reserve capacities of M0 and M1 are compared in Table III.

TABLE III
ECONOMIC EFFICIENCY EVALUATION IN PJM 5-BUS SYSTEM

| Methods | M0 | M1 |
|---|---|---|
| Total operational cost ($) | **1.679×10⁴** | 1.714×10⁴ |
| Energy cost ($) | **1.436×10⁴** | 1.386×10⁴ |
| Reserve cost ($) | **2.434×10³** | 3.290×10³ |
| Total up reserve capacity (MW) | **326.79** | 329.00 |
| Total down reserve capacity (MW) | **160.00** | 329.00 |
| Max transmission violation | **5.00%** | 4.89% |
| Max generation violation | **5.00%** | 4.95% |

The results show that while M1 is feasible, it will impose a higher cost on system operations since sufficient reserve capacity must be online to manage the uncertainty of the wind power. The proposed method allows wind generators to provide flexible scheduling by dispatching the wind curtailment to reduce the total operational cost. The total reserve capacities are significantly reduced by the proposed method, which means that the wind curtailment dispatch can relieve the wind uncertainty and provide a promising way to save the reserve resources.

### D. Results on IEEE 118-bus System

The IEEE 118-bus system is used to demonstrate the scalability of the proposed method. Four wind power plants are connected at buses 2, 34, 80, and 110. The forecasted wind power outputs and statistical moments of uncertainty are listed in Table IV. The scheduling results and computing times are given in Table V.

TABLE IV
WIND UNCERTAINTY MEAN AND STANDARD DEVIATION

|   | $W_{fc}$ (MW) | μ (MW) | σ (MW) |
|---|---|---|---|
| W1 | 200 | 0 | 30 |
| W2 | 200 | 0 | 40 |
| W3 | 200 | 0 | 40 |
| W4 | 200 | 0 | 60 |

TABLE V
ECONOMIC EFFICIENCY EVALUATION IN IEEE 118-BUS SYSTEM

| Methods | M0 | M1 |
|---|---|---|
| Total operational cost ($) | **1.535×10⁵** | 1.549×10⁵ |
| Energy cost ($) | **1.491×10⁵** | 1.486×10⁵ |
| Reserve cost ($) | **4.460×10³** | 6.351×10³ |
| Max transmission violation | **5.00%** | 5.00% |
| Max generation violation | **5.00%** | 5.02% |
| Computation time (s) | **5.6702** | 1.3472 |

Table IV demonstrates that considering wind curtailment dispatch reduces the total operational cost and the reserve cost. The energy cost is mildly raised because more conventional generator outputs are needed to overcome system-wide power unbalance issues caused by wind curtailment. The comparison results also show that the proposed method can deal with larger-scale systems and the computation time is acceptable. Hence, the proposed method provides a more economical and flexible solution to address wind integration into the transmission system.

## VI. CONCLUSIONS AND FUTURE WORK

This paper proposes a data-driven framework for CCO with wind curtailment dispatch. Specifically, the wind curtailment is modeled as a cap enforced on the wind power output and the GP surrogates are developed to describe the relationship between wind curtailment and the chance constraints. This paves the way of reformulating the CCO as a MI-SOCP problem. A correction strategy is also proposed to enhance the modeling accuracy by solving a convex LP problem. Comparisons with the recent CCO methods show that the proposed approach can strictly maintain the preset risk probability while achieving high economic performance no matter wind curtailment is used or not. The solutions obtained by the proposed approach can find the appropriate level of wind curtailment to balance the operational cost and security. The future work will be on extending the proposed approach to the stochastic market problem.

## APPENDIX A
### UNCERTAINTY PART OF POWER FLOWS AND RESERVE REQUIREMENTS

The transmission power flows $PF_l$ in (1c), (1d) can be formulated considering (1j) to (1l).

$$\begin{aligned} PF_l &= \sum_{i\in I} PTDF_{il}(P_i + W_i - D_i) \\ &= \sum_{i\in I} PTDF_{il}\left[P_{sc,i} - \beta_i \sum_{i\in I} \Delta W_i(wc_i) + W_{fv,i} + \Delta W_i(wc_i) - D_i\right] \\ &= \sum_{i\in I} PTDF_{il}\left[P_{sc,i} + W_{fv,i} - D\right] + \\ &\quad \sum_{i\in I}\left[(-\sum_{k\in I} PTDF_{kl}\beta_k + PTDF_{il})\Delta W_i(wc_i)\right] \end{aligned}$$ (19)

For each scenario $\xi^n$, the transmission power flows caused by wind uncertainty can be expressed as:

$$\Delta PF_l = \sum_{i\in I}\left[(-\sum_{k\in I} PTDF_{kl}\beta_k + PTDF_{il})\Delta W_i^n(wc_i)\right]. \quad (20)$$

Similarly, the reserve requirements caused by wind uncertainty can be formulated as follows:

$$\Delta R_i = -\beta_i \sum_{i\in I_W} \Delta W_i(wc_i). \quad (21)$$

## APPENDIX B
### DETAILED FORMULATION FOR LP PROBLEM

$$\min \sum_{i\in I}\left(c_i^1 \cdot (P_{sc,i} - \beta_i \sum_{i\in I_W} \mu_i(wc_i)) + c_i^2 \cdot (R_{up,i} + R_{dn,i})\right) \quad (22a)$$

s.t. Constraints (3b) and (3g)-(3i);

$$PF_{sc,l} + \sum_{i\in I_W} K_{il}\mu_i(wc_i) + \Psi^{-1}(1-\varepsilon)\Lambda_{PF,l} + \kappa_{\max,l}(\boldsymbol{wc}) \le F_l \quad (22b)$$

$$-PF_{sc,l} - \sum_{i \in I_w} K_{li}\mu_i(wc_i) + \Psi^{-1}(1-\varepsilon)\Lambda_{PF,l} + \kappa_{\min,l}(\boldsymbol{wc}) \leq F_l \quad (22c)$$

$$-\beta_i \sum_{i \in I} \mu_i(wc_i) + \Psi^{-1}(1-\varepsilon)\Lambda_{g,i} + \kappa_{up,i}(\boldsymbol{wc}) \leq R_{up,i} \quad (22d)$$

$$\beta_i \sum_{i \in I} \mu_i(wc_i) + \Psi^{-1}(1-\varepsilon)\Lambda_{g,l} + \kappa_{dn,i}(\boldsymbol{wc}) \leq R_{dn,i} \quad (22e)$$

$$\sqrt{\sum_{i \in I_w} K_{li}^2 \sigma_i(wc_i)^2} \leq \Lambda_{PF,l} \quad (22f)$$

$$\sqrt{\beta_i^2 \sum_{i \in I_w} \sigma_i(wc_i)^2} \leq \Lambda_{g,l}. \quad (22g)$$